\documentstyle[aps]{revtex}
\begin{document}
\def\baselinestretch{1.2}
%\hfill{{\tt draft}
\begin{center}
{\bf Mainz ``measurement" of the $E2/M1$ ratio in the $N-\Delta$ transition}
\end{center}

Beck {\it et al.} \cite{beck} have recently reported precise measurements
of differential cross sections and polarized photon asymmetries on the
reaction $\vec{\gamma}p$ $\rightarrow$ $p\pi^0$, using
tagged photons in the energy region 270 to 420 MeV, thus spanning the 
$\Delta$(1232) resonance. This augments the data
from the Brookhaven LEGS facility \cite{andy}.

Let us emphasize from the outset that the $E2/M1$ ratio in the
$N-\Delta$ transition \begin{it}is not directly measured\end{it}
by Beck {\it et al.},
despite the title of their paper. This is an {\it inferred} quantity
requiring theoretical modelling of the data.
Here, we take issue with some points of the
analysis reported by Beck {\it et al.} We show that our $E2/M1$
ratio, $R_{EM}$, extracted from the data of Beck {\it et al.} \cite{beck}
is {\it substantially} different
from what is obtained in Ref.~\cite{beck}: while Beck {\it et al.} obtain
this ratio to be -(2.5$\pm$0.2$\pm$0.2)\%, we get -(3.19$\pm$0.24)\%.
This difference is mostly due to the inaccuracy introduced by the use of
approximations in identifying $R$=$C_{\|}/(12A_{\|})$ with $R_{EM}$,
in Eqs. (7,8) of Ref.~\cite{beck}. We also emphasize that the systematic
error of $\pm$ 0.2\% for $R_{EM}$ estimated by Beck {\it et al.} due to
``... limited angular efficiency for detecting the recoil proton ...
and from ignoring the isospin 1/2 contributions", {\it does not}
include the error made by them in ignoring the $E_{1+}$ multipole in
$A_{\|}$.

We start with the coefficients characterizing the differential cross section,
assuming dominance of s- and p- waves:
\begin{eqnarray}
A_{\|} &=& |E_{0+}| ^2 + | 3E_{1+}-M_{1+}+M_{1-}| ^2 \; , \\
B_{\|} &=& 2{\rm Re} \left[ E_{0+}(3E_{1+}+M_{1+}-M_{1-}) ^* \right] 
\; , \\
C_{\|} &=& 12 {\rm Re}\left[ E_{1+}(M_{1+}-M_{1-}) ^* \right] \; ,
\end{eqnarray}
correcting an error in Eq.~(4) of Ref.~\cite{beck}. Key to the analysis of
Beck {\it et al.} is identifying $R$ with $R_{EM}$. This is imprecise
for the following reasons. First, this requires neglecting $M_{1-}$,
$E_{0+}$ and the isospin 1/2 components of $M_{1+}$ and $E_{1+}$ in
Eqs.~(1-3), and in addition neglecting $E_{1+}$ in Eq.~(1) altogether.
Second, equality of $R$ and $R_{EM}$ is not a good approximation even at
the K-matrix pole as implicitly assumed in Ref.~\cite{beck}. It gets far
worse, away from this pole. Finally, contrary to the
assertions of Ref.~\cite{beck}, Re$(M_{1+}-M_{1-})$ is {\it not} zero and
Im$M_{1+}$, Im$M_{1-}$ are {\it not} purely isospin 3/2, even at the K-matrix
pole. These effects need to be estimated in a model, as done by us below. 
We realize that some of these approximations
are unavoidable for Beck {\it et al.} in order to extract $R_{EM}$ from
the data, in absence of a model.
The best they can do is not to
neglect $E_{1+}$ in Eq.~(1), as we show below. 

We use our effective Lagrangian approach \cite{dmw} to analyze the Mainz
data set without making any of the above approximations, and retaining
partial waves beyond s and p. We get at the K-matrix pole, 
338.4$\pm$ 0.5 MeV, $M1$ = 282.5$\pm$ 1.3,
$E2$ = -9.00$\pm$0.66, both in units of 10$^{-3}$GeV$^{-1/2}$, and
$R_{EM}$ =
-(3.19$\pm$0.24)\%; at 340 MeV, we get $R_{EM}$ = -(3.09$\pm$0.24)\%.
The value of $R$ at 340 MeV is -(2.69$\pm$0.17)\%,
consistent with the result of
Ref.~\cite{beck}. The difference between $R$ and $R_{EM}$, given here,
is mainly due to the isospin 3/2 piece of the $E_{1+}$ in
Eq.~(1), neglected by Beck {\it et al.} This can be verified by using
{\it their} value of $R$ and correcting for the isopsin 3/2
piece of the $E_{1+}$ amplitude. This gives $R_{EM}$ $\approx$
-(2.9$\pm$0.23)\%, in agreement with our value.

A comparison between the LEGS \cite{andy} and the
Mainz\cite{beck} published data indicates no
significant discrepancy between $R_{EM}$ inferred from the
former data base \cite{dm} and the present Mainz result presented here.
 
We thank R. Beck for sharing the Mainz data and
R. Workman for useful discussions.
This research is supported by the U.S.~Dept.~of Energy.

R. M. Davidson and Nimai C. Mukhopadhyay \\
{\it
 Department of Physics, Applied Physics and Astronomy\\
 Rensselaer Polytechnic Institute,
 Troy, NY 12180-3590}

\end{document}